\documentclass[twolcolumn, superscriptaddress,pre, reprint]{revtex4-2}
\usepackage[latin9]{inputenc}
\usepackage{verbatim}
\usepackage{amsmath}
\usepackage{amssymb}
\usepackage[scr=rsfs]{mathalpha}
\usepackage{graphicx}

\usepackage{amsthm}
\usepackage{fancyhdr}
\usepackage{epsfig}
\usepackage{bbm}
\usepackage{subfigure}
\usepackage{float}
\usepackage{xcolor}
\usepackage[normalem]{ulem}
\usepackage{hyperref}
\usepackage{physics}
\usepackage{notes2bib}

\begin{document}
\title{
The Dance of Odd-Diffusive Particles: A Fourier Approach 
}

\author{Amelie Langer}
\affiliation{University of Augsburg, Institute of Physics, D-86159 Augsburg, Germany}

\author{Abhinav Sharma}
\affiliation{University of Augsburg, Institute of Physics, D-86159 Augsburg, Germany}
\affiliation{Leibniz-Institute for Polymer Research, Institute Theory of Polymers, D-01069 Dresden, Germany}

\author{Ralf Metzler}
\affiliation{University of Potsdam, Institute of Physics and Astronomy, D-14476 Potsdam, Germany}
\affiliation{Asia Pacific Centre for Theoretical Physics, KR-37673 Pohang, Republic of Korea} 

\author{Erik Kalz}
\email{erik.kalz@uni-potsdam.de}
\affiliation{University of Potsdam, Institute of Physics and Astronomy, D-14476 Potsdam, Germany}

\begin{abstract}
Odd-diffusive systems are characterized by transverse responses and exhibit 
unconventional behaviors in interacting systems. To address the dynamical 
interparticle rearrangements in a minimal system, we here exactly solve the 
problem of two hard disk-like interacting odd-diffusing particles. 
We calculate the probability density function (PDF) of the interacting 
particles in the Fourier-Laplace domain and find that oddness rotates all 
modes except the zeroth, resembling a ``mutual rolling'' of interacting odd 
particles. We show that only the first Fourier mode of the PDF, the 
polarization, enters the calculation of the force autocorrelation function 
(FACF) for generic systems with central-force interactions. An analysis of 
the polarization as a function of time reveals that the relative rotation angle between 
interacting particles overshoots before relaxation, thereby rationalizing the 
recently observed oscillating FACF in odd-diffusive systems. 
\end{abstract}
\maketitle

\section{Introduction}

The description of dissipative systems with an inherent broken time-reversal 
or parity symmetry has recently attracted considerable 
attention~\cite{hargus2021odd,kalz2022collisions, yasuda2022time, 
fruchart2023odd, hargus2024flux}. Relevant systems can be 
found in various domains of statistical physics, and include Brownian particles 
under the effect of Lorentz force \cite{chun2018emergence}, 
skyrmionic spin structures \cite{schick2024two} and active chiral
 particles \cite{hargus2021odd}. The transport 
 coefficients of these systems show a characteristic transverse response to 
 perturbations, which is encoded in antisymmetric off-diagonal elements in 
 transport tensors. These characteristic elements behave odd under the 
 transformation of the underlying (broken) symmetry, which serves as the 
 namesake for these \textit{odd} systems. Interestingly, a transverse response 
 does not necessarily imply an anisotropic description of the system. In fact, 
 in two spatial dimensions, odd transport coefficients represent the most 
 general description of an isotropic physical system~\cite{avron1998odd}. In 
 this work, we specifically consider odd-diffusive systems~\cite{hargus2021odd, 
 kalz2022collisions}, which in two spatial dimensions are characterized by a 
 diffusion tensor of the form
\begin{equation}
\label{eq:odd_diffusion_tensor}
    \mathbf{D} = D_0 (\mathbf{1} + \kappa \boldsymbol{\epsilon}),
\end{equation}
where $D_0$ is the bare diffusivity with physical diemensions 
$[D_0] = \mathrm{m}^2/\mathrm{s}$, and $\kappa$ is the characteristic 
odd-diffusion parameter with $[\kappa] = 1$, encoding the transverse response. 
$\mathbf{1}$ is the identity tensor and $\boldsymbol{\epsilon}$ the fully 
antisymmetric Levi-Civita symbol in two dimensions ($\epsilon_{xy}=
-\epsilon_{yx} =1$ and $\epsilon_{xx}=\epsilon_{yy} =0$). 

While the oddness parameter enters explicitly in the diffusion tensor 
in Eq.~\eqref{eq:odd_diffusion_tensor}, the mean-squared displacement of a 
freely diffusing particle is determined only by the symmetric part of the 
diffusion tensor. However, in a system of interacting particles, oddness 
qualitatively alters the diffusive behavior by affecting the 
self-diffusion~\cite{kalz2022collisions, kalz2024oscillatory}. 
The self-diffusion measures the dynamic of a tagged particle in a crowded 
system and explicitly accounts for interactions of 
particles~\cite{dhont1996introduction}. Independent of the microscopic 
details of the inter-particle interactions, the self-diffusion is usually 
reduced~\cite{batchelor1976brownian, felderhof1978diffusion, hanna1982self, 
abney1989self, zahn1997hydrodynamic, bembenek2000role, jones1979diffusion, 
medina1988long, lowen1993long,imhof1995long, thorneywork2015effect}. In 
odd-diffusive systems, however, it was recently shown that even purely 
repulsive interactions can enhance the self-diffusion~\cite{kalz2022collisions}. 
Via expressing the self-diffusion as a time-integral of the (interaction) 
force autocorrelation function (FACF)~\cite{hanna1981velocity}, these findings 
could further be related to the unusual microscopic particle rearrangements 
in odd-diffusive systems~\cite{kalz2024oscillatory}. 

For the overdamped equilibrium system under consideration here and consistent 
with the reduction of the self-diffusion~\cite{hanna1981velocity}, 
autocorrelation functions in general decay monotonically in 
time~\cite{feller1971introduction}. However, the enhancement of self-diffusion 
for odd systems can occur only if the FACF switches sign, i.e., becomes 
non-monotonic in time. This apparent contradiction could be resolved by 
recognizing that the time-evolution operator in an odd-diffusive system becomes 
non-Hermitian when $\kappa\neq 0$ in Eq.~\eqref{eq:odd_diffusion_tensor}, 
thereby breaking the monotonicity requirements on the 
FACF~\cite{kalz2024oscillatory}. Surprisingly, in this work, it was further 
observed that the FACF even oscillates in time. While this is consistent with 
the observed enhancement of the self-diffusion in odd systems, a physical 
interpretation of this phenomenon remains elusive.

In the present work, we show that the non-monotonic FACF originates in the 
unusual dynamics of interacting odd particles. When a pair of odd-diffusive 
particles interacts, their motion resembles that of two mutually rotating 
particles~\cite{kalz2022collisions}. This is despite the fact that the 
interaction potential is central-force, i.e., it acts along the vector that 
connects the centers of two particles. Our approach is based on the exact 
analytical derivation of the propagator of two interacting odd-diffusive 
particles. The joint probability density function (PDF) for the two particles 
separates into a center-of-mass PDF and a PDF of the relative coordinate, 
capturing the interplay of odd diffusion and interactions. We can express the 
relative PDF in a Fourier series to find that only certain modes enter the 
averages of observables, such as in the FACF. In particular, it is the 
polarization mode of the relative PDF, representing the positioning of the 
particles with respect to each other in time, which determines the full 
tensorial force autocorrelation behavior. By analyzing the polarization mode 
in time, we understand what originates the oscillating FACF. Interacting 
odd-diffusive particles mutually rotate further in time before they eventually 
relax. Figuratively they perform a dance, reminiscent of the classical step 
of the Viennese waltz.

The remainder of this work is organized as follows: In 
Section~\ref{sec:main_part} we set the problem of two interacting odd-diffusive 
particles, which is then exactly solved in Section~\ref{sec:analytic_solution}. 
The propagator can be put into the form of a Fourier series, of which the 
numerical analysis of the modes is presented in 
Section~\ref{sec:numerical_results}. In Section~\ref{sec:relevance_polarization} 
we show that only certain Fourier modes enter in averages of observabels, in 
particular into the FACF. Finally, in Section~\ref{sec:discussion} we summarize 
and give an extensive overview of systems, which can be subsumed under the 
terminology of odd diffusion. In Appendix~\ref{app:inner_solution} we present 
the detailed solution of the problem of interacting particles and 
Appendix~\ref{app:integral_relations} lists the relevant integral relations.

\section{Interacting odd-diffusive particles}
\label{sec:main_part}

We study the dynamics of two odd-diffusive particles at 
positions $\mathbf{x}_1$ and $\mathbf{x}_2$ in two spatial dimensions. 
The particles are assumed to interact with the potential energy $U$, 
which we assume to be of the hard-disk type
\begin{equation}
\label{eq:hard_interaction_potential}
U(\mathbf{x}_1, \mathbf{x}_2) = U(r) = \begin{cases} \infty, & r \leq 1\\
     0, & r > 1 \end{cases},
\end{equation}
where $r = |\mathbf{x}_1 - \mathbf{x}_2|/d$ is the rescaled relative 
distance between the particles and $d$ is the particle diameter.

The conditional joint PDF, i.e., the propagator for the particles to be found 
at positions $(\mathbf{x}_1, \mathbf{x}_2)$ at time $t$ given that they were 
at positions $(\mathbf{x}_{1,0}, \mathbf{x}_{2,0})$ at time $t_0$, $P(t) = 
P(\mathbf{x}_1, \mathbf{x}_2, t|\mathbf{x}_{1,0}, \mathbf{x}_{2,0},t_0)$, 
evolves according to the time-evolution equation 
\begin{equation}
\label{eq:fpe}
\begin{split}
\pdv{t} P(t) &= \nabla_1 \cdot \, [ \mathbf{D}\nabla_1 +  \boldsymbol{\mu} 
\nabla_1 U(\mathbf{x}_1, \mathbf{x}_2)]\,P(t) \\ 
    &\quad  + \nabla_2 \cdot\,[ \mathbf{D} \nabla_2 +  \boldsymbol{\mu} 
    \nabla_2 U(\mathbf{x}_1, \mathbf{x}_2)]\,P(t),
\end{split}
\end{equation}
where the initial condition is given as $P(t=t_0) = \delta(\mathbf{x}_1 - 
\mathbf{x}_{1,0})\, \delta(\mathbf{x}_2- \mathbf{x}_{2,0})$. In 
Eq.~\eqref{eq:fpe}, $\nabla_1, \nabla_2$ are the partial differential operators 
for the positions of the particles, and $\mathbf{D}$ is the odd-diffusion tensor 
of Eq.~\eqref{eq:odd_diffusion_tensor}. $\boldsymbol{\mu}$ is the mobility 
tensor, and we assume the fluctuation-dissipation relation (FDR) to hold 
\begin{equation}
\label{eq:tensorial_FDR}
\mathbf{D} = k_\mathrm{B}T \boldsymbol{\mu},
\end{equation}
where $k_\mathrm{B}$ is the Boltzmann constant and $T$ the temperature of the 
solvent. Note that even though Eq.~\eqref{eq:fpe} looks formally equivalent to a 
Fokker-Planck equation for the joint PDF $P(t)$, strictly spoken it is not, due 
to the antisymmetric (odd-diffusive) elements in the diffusion 
tensor~\cite{oksendal2003stochastic}. However,  based on the assumption of the 
FDR Eq.~\eqref{eq:fpe} resembles equilibrium dynamics with a unique 
steady-state solution~\cite{kalz2024oscillatory, doi1988theory}.

\subsection{Analytical solution}
\label{sec:analytic_solution}

We rescale space with the diameter of the particle $d$, 
$\mathbf{x}_i \to \mathbf{x}_i/d$, and time by the natural time-scale of 
diffusing the radial distance of a particle diameter $\tau_d = d^2/(2D_0)$, $t 
\to \tau = t/\tau_d$. Given the radial symmetry of the interaction potential 
$U(r)$, the time-evolution equation~\eqref{eq:fpe} can be written in terms 
of a center-of-mass coordinate $\mathbf{x}_\mathrm{c} = (\mathbf{x}_1 + 
\mathbf{x}_2)/2$ and a relative coordinate $\mathbf{x} = \mathbf{x}_1 - 
\mathbf{x}_2$ as
\begin{equation}
\label{eq:fpe_sperated}
\begin{split}
\pdv{\tau} P(\tau) &= \frac{1}{4} \nabla_{\mathbf{x}_\mathrm{c}}^2\,P(\tau) \\ 
    &\quad + \nabla_{\mathbf{x}} \cdot(\mathbf{1} + \kappa\boldsymbol{\epsilon})
     [\nabla_{\mathbf{x}} +  \beta\, \nabla_{\mathbf{x}} U(r)]\,P(\tau),
\end{split}
\end{equation}
where $\beta = 1/k_\mathrm{B}T$ and $\nabla_{\mathbf{x}_\mathrm{c}}, 
\nabla_{\mathbf{x}}$ are the partial differential operators corresponding to 
the center-of-mass and relative coordinates. Note that $(\mathbf{1} + 
\kappa\boldsymbol{\epsilon}) = \mathbf{D}/D_0$ represents the dimensionless 
odd-diffusion tensor of Eq.~\eqref{eq:odd_diffusion_tensor}.  As 
Eq.~\eqref{eq:fpe_sperated} decouples the coordinates, the propagator   
can be written as $P(\mathbf{x}_1, \mathbf{x}_2, \tau|\mathbf{x}_{2,0}, 
\mathbf{x}_{1,0},\tau_0) = p_c(\mathbf{x}_\mathrm{c}, 
\tau|\mathbf{x}_{\mathrm{c},0}, \tau_0)\, p(\mathbf{x}, \tau|\mathbf{x}_0, 
\tau_0)$ and the center-of-mass problem can be solved straight forwardly in the
form 
\begin{equation}
\label{eq:com_PDF}
p_c(\mathbf{x}_\mathrm{c}, \tau|\mathbf{x}_{\mathrm{c},0}) = \frac{1}{\pi \tau}\  
\mathrm{exp}\left(- \frac{|\mathbf{x}_\mathrm{c} - \mathbf{x}_{\mathrm{c},0}|^2}
{\tau}\right),
\end{equation}
where we have set $\tau_0 = 0$ as the underlying stochastic process is 
time-translational invariant~\cite{balakrishnan2021elements}.

The relative PDF can be put into the form of a (Cartesian) multipole 
expansion~\cite{degennes1995the, jackson2021classical, cates2013active, 
kalz2024field, te2020relations} which in polar coordinates $\mathbf{x} = 
(r, \varphi)$,  $\mathbf{x}_0 = (r_0, \varphi_0)$, is given as
\begin{align}
\label{eq:multipole_expansion}
    &p(\mathbf{x},\tau|\mathbf{x}_0) =\Theta(r-1)\bigg[\varrho(r, \tau|r_0) 
    + \boldsymbol{\sigma}(r, \tau|r_0) \cdot \mathbf{e}(\Delta\varphi) 
    \nonumber\\ 
    &\quad + \boldsymbol{Q}(r, \tau |r_0):\left(\mathbf{e}(\Delta\varphi) 
    \otimes \mathbf{e}(\Delta\varphi) - \frac{\mathbf{1}}{2}\right)+ 
    \ldots \bigg],
\end{align}
where $\mathbf{e}(\Delta\varphi) = (\cos(\Delta\varphi), 
\sin(\Delta\varphi))^\mathrm{T}$, $\Delta\varphi = \varphi-\varphi_0$ is the 
angular difference. $\mathbf{Q}\colon(\mathbf{e}\otimes \mathbf{e} - 
\mathbf{1}/2)= \sum_{\alpha, \beta=1}^2 Q_{\alpha\beta}(e_\beta e_\alpha - 
\delta_{\beta\alpha}/2)$ denotes the full contraction, where 
$\mathbf{e}\otimes \mathbf{e}$ is the outer product. We again set $\tau_0=0$ 
here. Note the multiplicative Heaviside function $\Theta(r-1)$ in 
Eq.~\eqref{eq:multipole_expansion}, which is defined as $\Theta(x) =1$, if 
$x>1$ and $\Theta(x) =0$ otherwise. This ensures the no-overlap condition of the 
hard-disk interaction potential in Eq.~\eqref{eq:hard_interaction_potential}. 
The explicit Cartesian multipole expansion for the relative PDF in 
Eq.~\eqref{eq:multipole_expansion} constitutes the first terms of a Fourier 
expansion for $p(\mathbf{x},\tau|\mathbf{x}_0)$ which reads
\begin{align}
    p(\mathbf{x},\tau|\mathbf{x}_0) = \frac{\Theta(r-1)}{2\pi} \bigg[& 
    a_0(r, \tau|r_0) \\ & + 2\sum_{n=1}^\infty \begin{pmatrix}a_n(r, \tau|r_0) 
        \\ b_n(r, \tau|r_0)\end{pmatrix} \cdot \mathbf{e}(n\, \Delta\varphi)
        \bigg]. \nonumber
\end{align}
Here $a_n$ and $b_n$ are the Fourier coefficients of order $n$, $n\in 
\mathbb{N}_0$. We introduce the notation $p = \Theta(r-1)\, [p_0 + 
\sum_{n=1}^\infty p_n]$, where $p_0=a_0/2\pi$ and $p_n = [a_n\, \cos
\left(n\, \Delta\varphi\right) + b_n\, \sin\left(n\, \Delta\varphi\right)]/\pi$, 
$n\geq 1$ for our subsequent shorthand notation of the Fourier modes. The 
Cartesian modes in Eq.~\eqref{eq:multipole_expansion} are thus connected to 
the Fourier modes as
\begin{equation}
\label{eq:mean_positional_PDF}
    \varrho(r, \tau|r_0) = \frac{1}{2 \mathrm{\pi}}\  a_0(r,\tau|r_0),
\end{equation}
which is the (scalar) mean positional PDF, 
\begin{equation}
\label{eq:mean_polarization_PDF}
    \boldsymbol{\sigma}(r, \tau|r_0) = \frac{1}{\mathrm{\pi}} \begin{pmatrix} 
        a_1(r,\tau|r_0) \\ b_1(r,\tau|r_0)\end{pmatrix},
\end{equation}
which is the (vectorial) polarization order PDF, and 
\begin{equation}
\label{eq:mean_nematic_PDF}
    \boldsymbol{Q}(r, \tau|r_0) = \frac{1}{\mathrm{\pi}} \begin{pmatrix} 
        a_2(r,\tau|r_0) & \phantom{-} b_2(r,\tau|r_0)\\ b_2(r,\tau|r_0) & 
        -a_2(r,\tau|r_0)\end{pmatrix},
\end{equation}
which is the (tensorial) nematic order PDF.

In Appendix~\ref{app:inner_solution} we provide the full solution to the 
relative problem, thereby following the original derivations in 
Refs.~\cite{hanna1982self, kalz2024oscillatory}, and show that the general 
Fourier coefficients $a_n, b_n$ are given by

\begin{widetext}
\begin{align}
\label{eq:fourier_coefficient_a}
a_n(r,\tau|r_0) &=\frac{\mathrm{e}^{-\frac{r_0^2+r^2}{4\tau}}}{2\tau}\ I_n 
\left(\frac{r_0 r}{2\tau}\right) - \mathscr{L}^{-1} \Bigg\{ k_n(r, s|r_0) 
\bigg[s\, K_n^\prime(\sqrt{s}) \, I_n^\prime(\sqrt{s}) + (n \kappa)^2 
K_n(\sqrt{s})\,I_n(\sqrt{s}) - \delta(r_0 - 1)\, \frac{\sqrt{s}\, 
K_n^\prime(\sqrt{s})}{2K_n(\sqrt{s})}\bigg] \Bigg\}, \\
\label{eq:fourier_coefficient_b}
b_n(r,\tau|r_0) &= - n\kappa\,\left(1 - \frac{\delta(r_0 - 1)}{2}\right) \, 
\mathscr{L}^{-1} \left\{k_n(r, s|r_0) \right\},
\end{align}
\end{widetext}
where we used the abbreviation 
\begin{equation}
    k_n(r, s|r_0) = \frac{K_n(r\sqrt{s})\, K_n(r_0\sqrt{s})}{(\sqrt{s}\, 
    K^\prime_n(\sqrt{s}))^2 + (n\kappa\, K_n(\sqrt{s}))^2}.
\end{equation}
Here $s$ denotes the (dimensionless) Laplace variable. The Laplace transform 
for a function $f(\tau)$ is defined as $\mathscr{L}\{f\}(s) = 
\int_0^\infty\mathrm{d}\tau\ \mathrm{exp}(-s\tau)f(\tau)$, using the rescaled 
time variable $\tau = t/\tau_d$. The prime denotes the derivative $g^\prime(a) = 
\mathrm{d}g(x)/\mathrm{d}x|_{x=a}$ and $I_n(x),\, K_n(x)$ are the modified 
Bessel functions of the first kind and second kind, 
respectively~\cite{abramowitz1968handbook}. As apparent from 
Eqs.~\eqref{eq:fourier_coefficient_a} and~\eqref{eq:fourier_coefficient_b}, 
the inverse Laplace transformations remain unfeasible at the moment, and we 
rely on established numerical Laplace inversion methods. 

We stress that the problem of solving for the full propagator of interacting 
particles in Eq.~\eqref{eq:fpe} can be formulated in the language of 
scattering theory~\cite{franosch2010persistent}, thus allowing to use 
well-developed tools from quantum field theory~\cite{zinn2021quantum}. Odd 
diffusion here adds the perspective of non-Hermicity with potentially new 
insights~\cite{kalz2024oscillatory}.

\subsection{Numerical results and Fourier modes}
\label{sec:numerical_results}

\begin{figure*}
\centering
\includegraphics[width=\textwidth,clip=true]{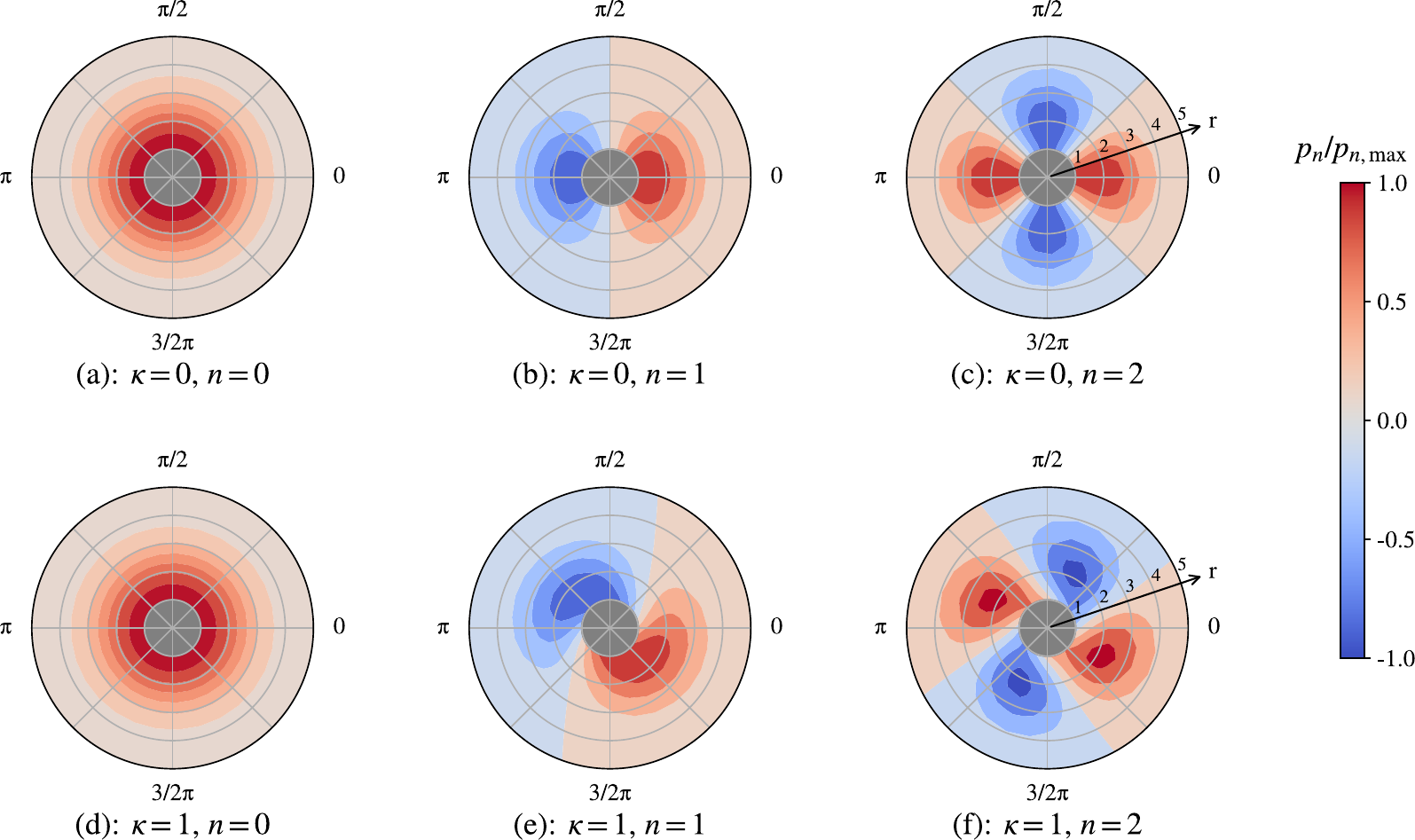}
\caption{Plots of the first three Fourier modes $p_0, p_1$ and $p_2$ of the 
relative PDF of interacting (odd-diffusive) particles $p(r, \tau|r_0) = 
\Theta(r-1)\, \sum_{n=0}^\infty\ p_n(r, \tau|r_0)$ at fixed time $\tau=1$. 
(a) - (c) show the modes for normal particles ($\kappa=0$) and (d) - (f) for 
odd-diffusive particles ($\kappa=1$). The initial positions are chosen as 
$(r_0, \varphi_0) = (1.01, 0)$, i.e. the particles are placed $1.01$ times 
their diameter along an arbitrarily defined $x$-axis. The zeroth-order mode 
in (a) and (d) equals the mean positional PDF for the relative coordinate $r$ 
(see Eq.~\eqref{eq:mean_positional_PDF}) which is unaffected by odd diffusion. 
The first order mode in (b) and (e), which corresponds to the (contracted) 
polarizational order PDF (see Eq.~\eqref{eq:mean_polarization_PDF}), and the 
second order mode in (c) and (f), which corresponds to the (contracted) 
nematic order PDF (see Eq.~\eqref{eq:mean_polarization_PDF}), as well as all 
higher order modes are affected by odd diffusion such that the modes are 
rotated in time compared to the $\kappa=0$ case, see also 
Fig.~\ref{fig:Polarisation}. Note that the gray circle in the middle of each 
plot is due to the excluded volume condition $\Theta(r-1)$ of the hard-disk 
interaction of the particles.}
\label{fig:Modes0to2}
\end{figure*}

\begin{figure*}
\centering
\includegraphics[width=\textwidth,clip=true]{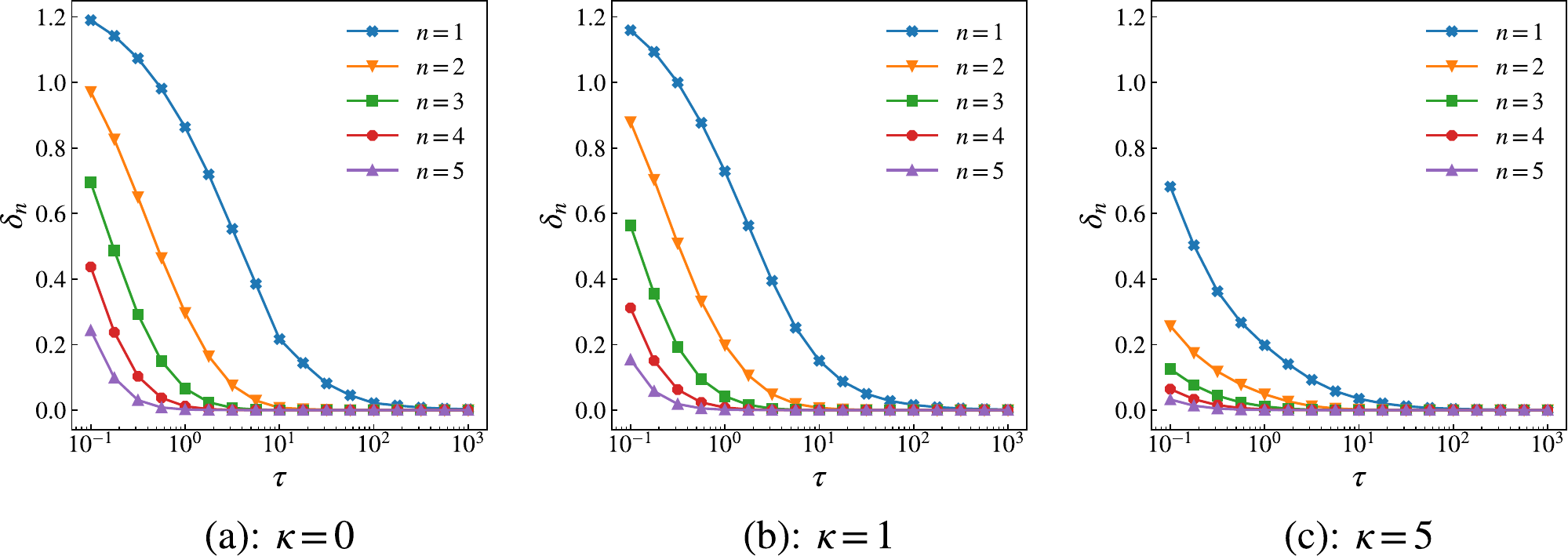}
\caption{Significance $\delta_n(\tau)$ of a mode $n$ as defined in 
Eq.~\eqref{eq:relDefDelta}, 
which measures the contribution of a Fourier mode of order $n\geq 1$, 
$p_n(\tau)$, to the full relative PDF $p(\tau) = \Theta(r-1)\, 
\sum_{n=0}^\infty\ p_n(\tau)$ as a function of time $\tau$ for odd-diffusion parameter 
$\kappa=0,1$ and $5$ in (a), (b) and (c), respectively. We see that higher 
order modes are consecutively ordered in their contribution to $p$ in time and 
become of comparable importance for $\tau\to 0$ as $p(\tau) \to p(0) \propto 
\delta(\mathbf{x} - \mathbf{x}_0)$. Comparing (a) with (c) we observe that 
$\kappa>0$ shifts the characteristic decay to shorter times. Thus, for dynamics 
where one is not interested in the $\tau\to 0$ limit, one can truncate the 
Taylor expansion at finite $n$.
\label{fig:RelevanceModes}}
\end{figure*}

\begin{figure*}
\centering
\includegraphics[width=\textwidth,clip=true]{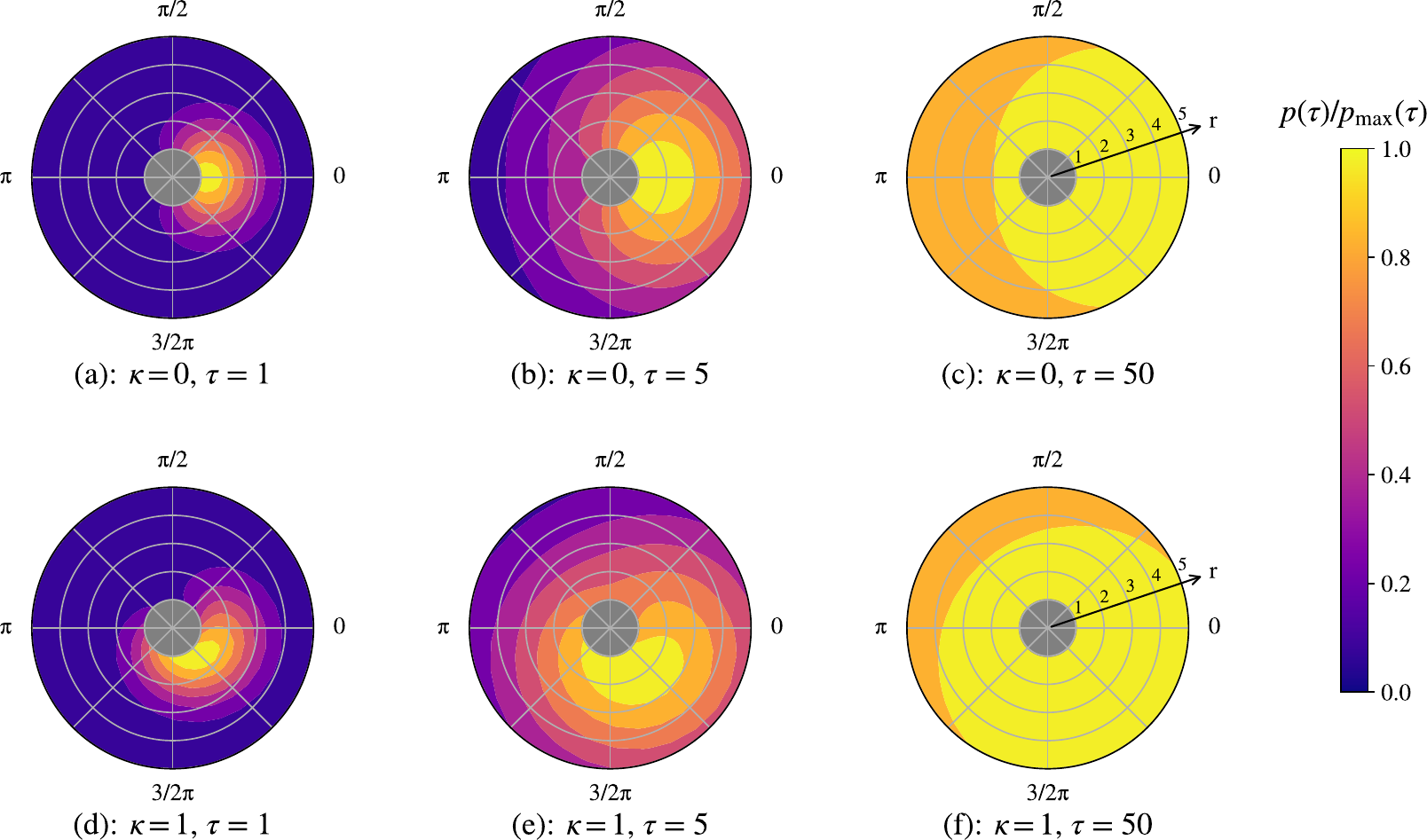} 
\caption{PDF $p(r,\varphi,\tau|r_0,\varphi_0)$ for the relative 
coordinate of two hard interacting particles. (a) to (c) show normal particles 
($\kappa = 0$) and (d) to (f) show odd-diffusive particles ($\kappa = 1$) at 
times $\tau=1,5$ and $50$. The initial positions are chosen as 
$(r_0, \varphi_0) = (1.01, 0)$, i.e. the particles are placed $1.01$ times 
their diameter along an arbitrarily defined $x$-axis. The Fourier expansion 
of $p(\tau)$ in Eq.~\eqref{eq:multipole_expansion} is truncated at $n=10$ in 
the numerical evaluation for the plots, see also Fig.~\ref{fig:RelevanceModes} 
for a rationale. (d) to (f) 
show the ``mutual rolling effect'', i.e. that odd-diffusive particles rotate 
around each other after a collision~\cite{kalz2022collisions}, which is in 
contrast to a symmetric back-reflection for normal particles, visible in  
(a) - (c). Note the grey circle in the middle of each plot which is due to the 
excluded volume condition $\Theta(r-1)$ of the hard-disk interaction of the 
particles.}
\label{fig:kappa0and1}
\end{figure*}

In Fig.~\ref{fig:Modes0to2} we compare the positional mode $p_0 = \Theta(r-1)\, 
\varrho$, the polarization mode $p_1 = \Theta(r-1)\, [\boldsymbol{\sigma} 
\cdot \mathbf{e}]$, and the nematic mode $p_2 = \Theta(r-1)\, [\mathbf{Q} 
\colon (\mathbf{e}\otimes \mathbf{e} - \mathbf{1}/2)]$ for a normal 
($\kappa=0$) and an odd-diffusive ($\kappa=1$) system of interacting particles. 
The modes are evaluated at fixed time $\tau=1$ and for $(r_0, \varphi_0) = 
(1.01, 0)$, i.e. particles are initially placed at a distance of $1.01$ times 
their diameter along the (arbitrarily) chosen $x$-axis.

We observe that $\kappa\neq 0$ does not affect the interacting particles' 
mean positional distribution $p_0$, but rotates higher order modes in 
comparison to $\kappa=0$. We can understand this by observing from 
Eq.~\eqref{eq:fourier_coefficient_b} that $b_n \propto n\kappa$, such that 
$p_0$ is not affected by $\kappa\neq 0$---but for higher order modes $b_n$ 
contributes for an odd-diffusive system. It is of interest to analyse the 
contribution of the higher-order modes to the time evolution of the relative 
PDF $p$. We define a measure of the significance of a mode of order $n \geq 1$ 
as
\begin{equation}\label{eq:relDefDelta}
\delta_n(\tau) = \frac{\int \mathrm{d}\mathbf{x}\ |p_n(\mathbf{x},\tau|
\mathbf{x}_0)|}{ \int \mathrm{d}\mathbf{x}\ p_0(\mathbf{x},\tau|\mathbf{x}_0)}. 
\end{equation}
We take the absolute value $|p_n|$ in the definition of $\delta_n(\tau)$, 
as the full space integral for the polarization, nematic and all higher order 
modes vanishes due to the orthogonality of the harmonic functions, 
$\int\mathrm{d}\mathbf{x}\ p_{n} = 0$, $n\geq 1$. In contrast, 
$\int\mathrm{d}\mathbf{x}\ p_0 =1$ (see Appendix~\ref{app:integral_relations}), 
which we only include in Eq.~\eqref{eq:relDefDelta} to avoid numerical 
discretization errors.

Fig.~\ref{fig:RelevanceModes} shows $\delta_n(\tau)$ for $n=1,\ldots,5$ and for 
different values of the odd-diffusion parameter $\kappa =0, 1, 5$. When 
considering $\delta_n$ as a measure for the relevance of the $n$th order mode, 
we observe that the modes are ordered consecutively in their contribution to 
the relative PDF $p$ and that considering higher order modes becomes important 
for $\tau \to 0$ as $p(\tau) \to p(0) \propto \delta(\mathbf{x} - \mathbf{x}_0)$, 
but higher order modes become less important for $\tau \gg 0$. As can be seen 
in Figs.~\ref{fig:RelevanceModes}(b), (c), $\kappa > 0$ shifts the decay of 
$\delta_n(\tau)$ to even shorter times. 

In Fig.~\ref{fig:kappa0and1} we plot the full relative PDF $p(\tau)$ at times 
$\tau =1, 5, 50$ for a normal ($\kappa=0$) and an odd-diffusive ($\kappa=1$) 
system. Again we choose the initial condition to be $(r_0, \varphi_0) = 
(1.01, 0)$ and, based on the observation in Fig.~\ref{fig:RelevanceModes}, we 
truncated the Fourier series at $n=10$. Comparing the normal and 
odd-diffusive relative PDF, we observe that the $\kappa$-induced rotation of 
every but the zeroth order Fourier mode persists into the full PDF. For 
$\kappa=0$, $b_n=0$ for all modes, and the Fourier representation of the 
relative PDF therefore only contains cosine terms. The relative PDF of a normal 
particle thus is symmetric around $\Delta\varphi=0, \pi$ for all times. This 
symmetry implies that particles encounter the space from both sides after a 
collision with equal likelihood. However, this does not hold for odd-diffusive 
particles. $\kappa\neq 0$ introduces a handedness in the diffusive exploration 
of space. For interacting odd-diffusive particles, the additional rotational 
probability flux introduced via Eq.~\eqref{eq:odd_diffusion_tensor} results in 
a preferred direction after a collision depending on the sign of the 
odd-diffusion parameter $\kappa$. This effect was recently observed from 
Brownian dynamics simulations and termed as 
``mutual rolling''~\cite{kalz2022collisions}. The interaction-induced 
symmetry breaking has far-reaching consequences for observable transport 
coefficients such as the self-diffusion coefficient. In odd-diffusive systems, 
even though resembling equilibrium overdamped dynamics, the self-diffusion can 
be enhanced by interactions instead of being reduced as for a normal 
system~\cite{kalz2022collisions}. Even though, seemingly contradicting 
equilibrium statistical mechanics theorems~\cite{feller1971introduction, 
leitmann2017time, caraglio2022analytic}, the interaction-enhanced self-diffusion 
could recently be rationalized by observing that the time-evolution in 
odd-diffusive systems (see Eq.~\eqref{eq:fpe}) becomes non-Hermitian for 
finite $\kappa$~\cite{kalz2024oscillatory}.

\subsection{Relevance of the polarization mode for the force autocorrelation}
\label{sec:relevance_polarization}

The force autocorrelation tensor (FACT) $\mathbf{C}_F(\tau)$ encodes the most 
detailed microscopic information in an interacting system. Via a 
Taylor-Green-Kubo relation~\cite{taylor1922diffusion, green1954markoff, 
kubo1957statistical} it encodes the particle-particle interaction effects in 
the self-diffusion. For a stationary system, $\mathbf{C}_F(\tau)$ is defined as
\begin{align}
\label{eq:full_FACF}
    \mathbf{C}_F(\tau) &= \langle \mathbf{F}(\vec{\mathbf{x}}) \otimes 
    \mathbf{F}(\vec{\mathbf{x}}_0)\rangle \\
    &= \int\mathrm{d}\vec{\mathbf{x}} \int\mathrm{d}\vec{\mathbf{x}}_{0}\ 
    \mathbf{F}(\vec{\mathbf{x}}) \otimes \mathbf{F}(\vec{\mathbf{x}}_0)\, 
    P_N(\vec{\mathbf{x}}, \tau, \vec{\mathbf{x}}_{0}, \tau_0)\nonumber,
\end{align}
where $\vec{\mathbf{x}} = \{\mathbf{x}_1, \ldots, \mathbf{x}_N\}$ and similarly 
$\vec{\mathbf{x}}_{0}$ for a system of, in general, $N$ particles. $P_N$ is the 
$N$-particle joint PDF, which can be rewritten as $P_N(\vec{\mathbf{x}}, \tau, 
\vec{\mathbf{x}}_{0}, \tau_0) = P_N(\vec{\mathbf{x}}, \tau| 
\vec{\mathbf{x}}_{0})\, P_\mathrm{eq}(\vec{\mathbf{x}}_{0}, \tau_0)$, 
assuming the particles where in equilibrium at time $\tau_0$, which we take 
again to be $0$. The force on the tagged particle (particle one) is 
$\mathbf{F}(\vec{\mathbf{x}}) = - \nabla_1 U_N (\vec{\mathbf{x}})$, and we 
assume a pairwise additive and radially symmetric potential 
$U_N (\vec{\mathbf{x}}) = \sum_{i,j =1}^{N} U(r_{ij})/2$, where 
$r_{ij} = |\mathbf{x}_i - \mathbf{x}_j|, i\neq j$. In the dilute limit, we can 
safely assume that only two-body correlations are important and thus ignore 
correlations between the untagged particles.
The equilibrium PDF can be approximated as $P_\mathrm{eq}(\vec{\mathbf{x}}_{0}) 
= 1/(VZ_N) \, \prod_{i=2}^N \mathrm{exp}(-\beta\, U(r_{1i,0})/2)$, where 
$V \subset \mathbb{R}^2$ is the bounded space of diffusion and 
$Z_N=\int\mathrm{d}\vec{\mathbf{x}}_0\ \prod_{i=2}^N 
\mathrm{exp}(-\beta\, U(r_{1i,0})/2)$ is the $N$-particle partition function. 
The conditional PDF, i.e., the $N$-particle propagator can be approximated as 
$P_N(\vec{\mathbf{x}}, \tau| \vec{\mathbf{x}}_{0})= (1/V)\, \prod_{i=2}^N 
p(\mathbf{x}_{1i}, \tau| \mathbf{x}_{1i, 0})$, where 
$p(\mathbf{x}_{1i}, \tau| \mathbf{x}_{1i, 0})$ is the PDF of the relative 
coordinate $\mathbf{x}_{1i} = \mathbf{x}_1 - \mathbf{x}_i$, similar to 
Eq.~\eqref{eq:multipole_expansion} but for a generic two-body interaction 
potential $U$. Following these approximations, all but one particle 
(particle two) can be integrated out in Eq.~\eqref{eq:full_FACF} and we denote 
as before, $\mathbf{x} = \mathbf{x}_{1} - \mathbf{x}_2$. The central interaction 
force between the particles can be written as $\mathbf{F}(\mathbf{x}) = 
F(r)\, \mathbf{e}(\varphi)$, where $\mathbf{e}(\varphi) = (\cos(\varphi), 
\sin(\varphi))^\mathrm{T}$, as before, and coincides with the radial unit vector. 
In the dilute limit, the FACT of Eq.~\eqref{eq:full_FACF} thus becomes 

\begin{align}
\label{eq:FACF_outer_product}
    \mathbf{C}_F(\tau) &= \frac{N-1}{V} \int\mathrm{d}\mathbf{x} 
    \int\mathrm{d}\mathbf{x}_{0}\  F(r)\,  F(r_0)\, \frac{\mathrm{e}^{-\beta 
    U(r_0)}}{Z_2} \nonumber \\
    & \quad  \times p(\mathbf{x}, \tau| \mathbf{x}_0)\, \mathbf{e}(\varphi) 
    \otimes \mathbf{e}(\varphi_0) \\
    \label{eq:FACF_polarization}
    &= \frac{N-1}{V} \pi \int\mathrm{d}r \int\mathrm{d}r_{0}\ F(r)\, F(r_0)\, 
    \frac{\mathrm{e}^{-\beta U(r_0)}}{Z_2} \nonumber \\
    & \quad \times [a_1(r, \tau|r_0)\, \mathbf{1} - b_1(r, \tau|r_0)\, 
    \boldsymbol{\epsilon}].
\end{align}
Here we used the orthogonality of the Fourier modes in the relative PDF 
$p(\tau)$ (analogously to Eq.~\eqref{eq:multipole_expansion}) from 
Eq.~\eqref{eq:FACF_outer_product} to Eq.~\eqref{eq:FACF_polarization} to find 
that only the polarization mode $\boldsymbol{\sigma}(r, \tau|r_0) = 
(a_1(r, \tau|r_0), b_1(r, \tau|r_0))^\mathrm{T}/\pi$ (analogously to 
Eq.~\eqref{eq:mean_polarization_PDF}) contributes to the FACT. Note that 
relations analogous to Eqs.~\eqref{eq:multipole_expansion} and 
\eqref{eq:mean_polarization_PDF} also hold in a system with generic radially 
symmetric interaction potential $U(r)$, specifically also for forces with 
transverse, odd components~\cite{shinde2022strongly, ghimenti2023sampling, 
schick2024two, batton2024microscopic}.

We can specify Eq.~\eqref{eq:FACF_polarization} for the hard-disk interaction 
potential of Eq.~\eqref{eq:hard_interaction_potential} by observing 
that $P_\mathrm{eq}(\mathbf{x}_{0}) = \Theta(r_0-1)/V^2$ and that we can 
rewrite the singular interaction force via the trick $\beta \Theta(r-1)\, 
\mathbf{F}(r) = \delta(r-1)\, \mathbf{e}(\varphi)$ \cite{trick}, obeying the same generic 
form of a central force as before. We can perform the radial integral of 
Eq.~\eqref{eq:FACF_polarization} and find that the FACT of a hard-disk system 
in the dilute limit is given by $\mathbf{C}_F(\tau) = \beta^{-2}\phi\, 
[a_1(1, \tau|1)\, \mathbf{1} - b_1(1, \tau|1)\, \boldsymbol{\epsilon}]$, 
where $\phi=(\pi d^2/4)\, (N/V)$ is the area fraction in dimensional 
form~\cite{kalz2024oscillatory, hanna1981velocity}. Thus we understand that 
the behavior of the polarization mode $\boldsymbol{\sigma}(\tau)$ in time 
governs the behavior of the FACT for interacting systems, in particular for a 
hard-disk interaction. Note specifically here, that the off-diagonal correlation 
($\propto  b_1\, \boldsymbol{\epsilon}$) is directly proportional to 
$\kappa$, and therefore might serve as a characteristic to odd 
diffusion~\cite{yasuda2022time, kalz2024oscillatory}.

\begin{figure*}[t]
\centering
\includegraphics[width=\textwidth,clip=true]{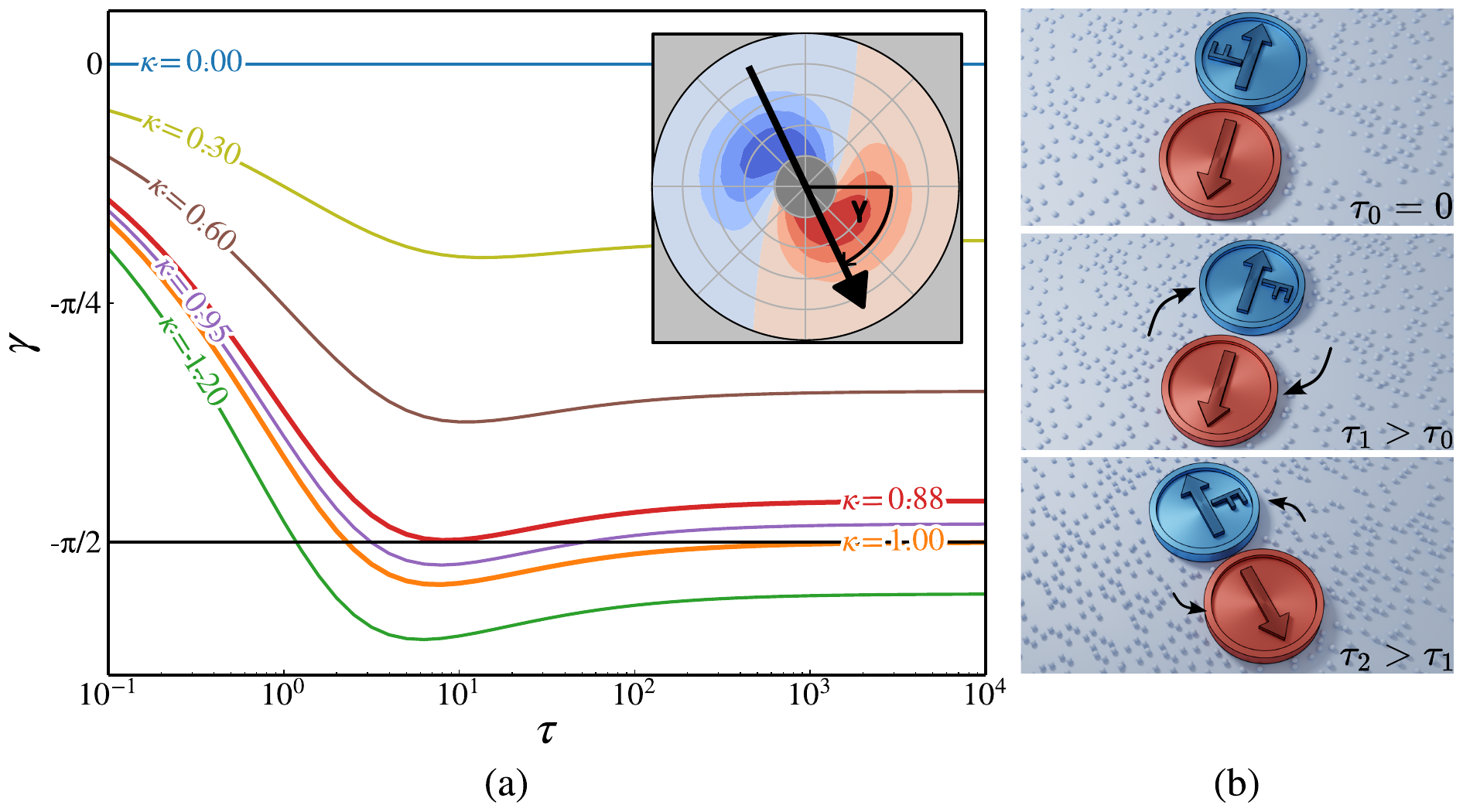}   
\caption{(a) Angle $\gamma(\tau)$ between the polarization mode 
$\boldsymbol{\sigma}(r=1, \tau|r_0=1)$, Eq.~\eqref{eq:mean_polarization_PDF}, 
and its initial direction as function of time $\tau$ for different odd-diffusion 
parameters $\kappa$, see Eq.~\eqref{eq:polarization_vector} and the inset. The 
polarization mode constitutes the vectorial mode of the relative PDF $p(\tau)$ 
for interacting particles and represents relative particle rearrangements. In a 
normal system ($\kappa=0$), $\gamma(\tau)=0$ and is constant in time, whereas 
it develops an extremum $\gamma_\mathrm{ext}$ different from its asymptotic 
steady state value $\gamma_\infty$ for $\kappa \neq 0$ in time. We observe that
$|\gamma_\mathrm{ext}| < |\gamma_\infty|$, indicating that particles on average 
rotate further in time than they finally relax to. Specifically for $\kappa 
\in (0.88, 1.0)$, $\gamma_\mathrm{ext} < -\pi/2$, but $\gamma_\infty > -\pi/2$ 
which quantitatively aligns with the oscillating force autocorrelation function 
previously observed for this regime~\cite{kalz2024oscillatory}. (b) The sketch 
qualitatively illustrates the scenario, that odd-diffusive particles with 
$0.88 < \kappa < 1$ collide at $\tau_0$ with each other along the (arbitrary) 
$x$-axis and rotate for $\tau_1>\tau_0$ to $\gamma_\mathrm{ext} < -\pi/2$, and 
eventually relax at $\tau_2 >\tau_1$ to  $\gamma_\infty > -\pi/2$.}
\label{fig:Polarisation}
\end{figure*}

To further analyze the polarization mode in the hard system, we define 
\begin{equation}
\label{eq:polarization_vector}
    \gamma(\tau) = \arctan\left(\frac{b_1(1,\tau|1)}{a_1(1,\tau|1)}\right),
\end{equation} 
as the mean angle between the polarization vector $\boldsymbol{\sigma}(1, \tau|1)$
as a function of time and its initial direction, which defines the $x$-axis in the system, 
see also inset in Fig.~\ref{fig:Polarisation}(a). We plot $\gamma(\tau)$ for 
different values of $\kappa$ as a function of time in 
Fig.~\ref{fig:Polarisation}(a) and use the $\mathrm{arctan2}(\cdot)$-function 
for numerical evaluation to obtain angles within the interval $(-\pi, \pi)$. 
We observe that for $\kappa=0$ the initial direction of the polarization does 
not change with time and stays constant. Recalling the polarization mode in 
Fig.~\ref{fig:Modes0to2}(b), the location of the extrema thus remain unchanged, 
meaning that the particles are symmetrically back-reflected from the center 
of the collision. For $\kappa\neq 0$, however, $\gamma(\tau)$ changes in time 
(see again Fig.~\ref{fig:Modes0to2}(e)). The mean-direction of the polarization 
develops an extremum $\gamma_\mathrm{ext}$ and relaxes to a constant, non-zero 
value $\gamma_\infty$ for long times. For $\kappa>0.88$,  
$\gamma_\mathrm{ext}< -\pi/2$ and for $\kappa>1$ even $\gamma_\infty < -\pi/2$. 
The interval $\kappa \in(0.88, 1)$ thereby aligns with the odd-diffusion 
interval, for which we reported an oscillating FACF 
earlier~\cite{kalz2024oscillatory}. These sign-switches in the FACF coincide 
with what we observe now for the direction of $\boldsymbol{\sigma}(\tau)$, see 
also inset in Fig.~\ref{fig:Polarisation}(a): $\gamma(\tau)$ crosses $-\pi/2$ 
for some time (negative FACF) and eventually relaxes to angles larger than 
$-\pi/2$ again (positive FACF).

We interpret this behavior as interacting odd-diffusive particles, which 
for $|\kappa|>0.88$ rotate more than $\pi/2$ but relax to a steady state with 
a relative angle $|\gamma_\infty| < \pi/2$ as long as $|\kappa|<1$; see 
Fig.~\ref{fig:Polarisation}(b) for a sketch. This behavior also rationalizes 
the oscillating FACF in the same interval, as $\langle\mathbf{F}(\tau)\cdot 
\mathbf{F}(0)\rangle = \langle F(\tau)\, F(0)\, \cos(\pi/2)\rangle = 0$ here 
and the particles oscillate around that angle. However, a steady state angle 
$|\gamma_\infty|$ which is smaller than the maximal angle 
$|\gamma_\mathrm{ext}|$ is a generic observation from 
Fig.~\ref{fig:Polarisation}(a). On average particles always rotate further in 
time, as they finally relax. Note that for $\kappa\to-\kappa$, we have that 
$\gamma(\tau) \to - \gamma(\tau)$, as $a_n$ is even and $b_n$ is an odd function 
of $\kappa$.  Thus, $\kappa<0$ results in the same phenomenology for the 
relaxation of $\gamma(\tau)$ and thus for the FACF. Further, if $|\kappa| \to 
\infty$, we find that $|\gamma_\mathrm{ext}| \to |\gamma_\infty| \to \pi$, 
which means, that at most odd particles exchange positions but do not rotate 
any further.

\section{Discussion}
\label{sec:discussion}

We here presented the exact analytical solution and numerical evaluation of two 
hard disk-like interacting odd-diffusing particles in two spatial dimensions. 
Odd diffusion thereby is characterized by antisymmetric elements $\propto \kappa$
in the diffusion tensor $\mathbf{D} = D_0 (\mathbf{1} + \kappa
\boldsymbol{\epsilon}$).
Our analysis showed that the two-particle problem separates into a 
center-of-mass and a relative coordinate problem, of which the first can be 
solved straightforwardly, and the latter incorporates the nontrivial effects 
of interactions and odd diffusion. The relative PDF can be written as a Fourier 
series, and we observe that oddness rotates all apart from the zeroth order mode 
in time, which represents the unaffected positional distribution of the relative 
coordinate. The modes are consecutively ordered in their contribution to the 
relative PDF, and when summed up show a rotated form of the PDF. This effect 
is a characteristic of odd diffusion and has been termed as 
``mutual rolling'' 
earlier~\cite{kalz2022diffusion} as the particles rotate around each other 
while interacting, in contrast to normal diffusive ($\kappa=0$) particles, 
which are symmetrically back-reflected when interacting.

The representation of the relative PDF in its Fourier modes becomes useful in 
the analysis of microscopic correlation functions, specifically the FACF. For 
any central interaction force, only the polarization mode of the relative PDF 
determines the correlation function. We conjecture here that a similar 
phenomenology holds for other (radially symmetric) observables of arbitrary 
tensorial order in an interacting system; the corresponding mode of the 
relative PDF might determine the expectation value and even the correlation 
function. We used the analytical access to the polarization mode, to understand 
the average configurations of the hard-disk-like interacting odd-diffusive 
particles. The maximal relative rotation angle overshoots the final relaxation 
on average, which quantitatively aligns with an oscillating FACF, recently 
reported for these systems~\cite{kalz2024oscillatory}.

Our work shines light on the novel way of particle interactions in odd-diffusive 
systems. Thereby it might serve as a reference case for interactions in the 
various odd-diffusive systems such as in equilibrium, e.g., Brownian particle 
under Lorentz force~\cite{lemons1999brownian, czopnik2001brownian, 
hayakawa2005langevin, simoes2005kramers, jimenez2006brownian, hou2009brownian, 
chun2018emergence, vuijk2019anomalous, chun2019effect, abdoli2020nondiffusive, 
abdoli2020correlations, park2021thermodynamic, abdoli2022tunable, 
kalz2022diffusion}, with a long-lasting history in statistical 
mechanics~\cite{taylor1961diffusion, kurcsunoglu1962brownian, karmeshu1974brownian}, 
or skyrmionic spin structures~\cite{schutte2014inertia, troncoso2014brownian, 
gruber2023300}. Here numerical studies recently reproduced the interaction-enhanced 
self-diffusion~\cite{schick2024two}, which originates in the mutual rolling 
effect~\cite{kalz2022collisions}. But there exist also non-equilibrium 
odd-diffusive systems, such as systems under shear~\cite{dieball2022coarse, 
dieball2022mathematical}, or active chiral particles~\cite{hargus2021odd, 
reichhardt2019active, muzzeddu2022active, poggioli2023odd, batton2024microscopic, 
chan2024chiral, kalz2024field, siebers2024collective}. In general, for systems 
that break time-reversal symmetry, the relevance of off-diagonal correlation 
functions for transport properties has attracted considerable interest 
recently~\cite{pavliotis2010asymptotic, hargus2021odd, yasuda2022time, 
vega2022diffusive, poggioli2023odd, batton2024microscopic, hargus2024flux}. 
Odd diffusion further might serve as the unifying terminology for systems 
showing transverse responses such as systems with Magnus 
forces~\cite{brown2018effect, reichhardt2022active, cao2023memory}, Coriolis 
forces~\cite{welander1966note, brandenburg2009turbulent}, in complex (porous) 
environments~\cite{koch1987symmetry, auriault2010asymmetry, marbach2018transport}, 
or to describe non-conservative force fields which are found in optical tweezer 
experiments~\cite{wu2009direct, sukhov2017non, mangeat2019role, 
volpe2023roadmap_short}. Systems with an artificial transverse interaction 
component~\cite{ghimenti2023sampling, ghimenti2024irreversible} recently showed 
that odd interactions enhance the sampling of configurations in dense systems, 
originating again in the unique particle rearrangements in odd-diffusive 
systems. Transverse forces further have been the subject of interest in so-called 
linear-diffusive systems~\cite{kwon2005structure, turitsyn2007statistics, 
kwon2011nonequilibrium, noh2013multiple, du2023dynamical}. Odd diffusion is 
finally also relevant for strongly rotating~\cite{kahlert2012magnetizing, 
hartmann2019self} or magnetized plasmas, as it can be found, e.g. in the realm 
of astrophysics, where antisymmetric transport is relevant to describe the 
movement of energetic particles moving through magnetized plasmas such as the 
interplanetary and interstellar medium~\cite{shalchi2011applicability, 
effenberger2012generalized, shalchi2020perpendicular}. 

\begin{acknowledgements} \textit{Acknowledgements.}
The authors acknowledge the help of Timo J. Doerries in the creation of 
Fig.~\ref{fig:Polarisation}(b). A. S.,  R. M., and E. K. acknowledge support 
by the Deutsche Forschungsgemeinschaft (grants No. SH 1275/5-1,  ME 1535/16-1 
and SPP 2332 - 492009952).
\end{acknowledgements}

\appendix
\section{Analytical solution of the relative problem}
\label{app:inner_solution}

This Appendix closely follows Ref.~\cite{kalz2024oscillatory} in its 
presentation of the analytical solution to the relative problem, which itself 
adapted the work of Hanna, Hess and Klein~\cite{hanna1982self} to odd-diffusive 
systems. 

The equation for the relative PDF $p(\tau) = p(\mathbf{x}, \tau|\mathbf{x}_0)$, 
where $\mathbf{x} = \mathbf{x}_1 - \mathbf{x}_2$, follows from 
Eq.~\eqref{eq:fpe_sperated} as
\begin{align}
\pdv{\tau} p(\tau) = \nabla_{\mathbf{x}} \cdot(\mathbf{1} + \kappa
\boldsymbol{\epsilon})[\nabla_{\mathbf{x}} +  \beta\, \nabla_{\mathbf{x}} 
U(r)]\,p(\tau),
\end{align}
and reads in (rescaled) polar coordinates $\mathbf{x} = (r, \varphi)$ as 
\begin{equation} 
\label{eq:EqRelCoord}
\begin{split}
\pdv{\tau} p(\tau) = \frac{1}{r^2}  \left( r\pdv{r} r\pdv{r} + r \pdv{r} \, 
r\beta \, \pdv{U(r)} {r} \right. \\
 \left.\quad + \pdv[2]{}{\varphi} - \kappa \,r \beta \, \pdv{U(r)}{r} 
 \pdv{\varphi} \right) \, p(\tau).
\end{split}
\end{equation}
Note that space is rescaled with the diameter of the particle $d$, 
$\mathbf{x} \to \mathbf{x}/d$, and time is rescaled by the natural 
time-scale of diffusing the radial distance of a particle diameter $\tau_d = 
d^2/(2D_0)$, $t \to \tau = t/\tau_d$. The initial condition $p(\tau=0) = 
\mathrm{\delta}(\mathbf{x} - \mathbf{x}_0)\, \mathrm{\Theta}(r-1)$ in 
polar coordinates becomes
\begin{equation}\label{eq:initial_cond_inner_p}
p(\tau=0) = \frac{\mathrm{\delta}(r - r_0)}{r_0}\, \mathrm{\delta}(\varphi - 
\varphi_0)\,\mathrm{\Theta}(r-1). 
\end{equation} 
The angular part of the initial condition can be expanded into a Fourier 
series $\mathrm{\delta}(\varphi - \varphi_0) = \sum_{n=-\infty}^\infty 
\mathrm{exp}(\mathrm{i}n (\varphi - \varphi_0)) /2\pi$, which, following 
Refs.~\cite{hanna1982self, kalz2024oscillatory}, we use as an ansatz to solve 
for $p(\tau)$ as
\begin{equation}
\label{eq:ansatz}
p(\mathbf{x},\tau|\mathbf{x}_0)  = \frac{\mathrm{\Theta}(r - 1)}{2\pi} 
\sum_{n=-\infty}^{\infty} R_n(r,\tau|r_0)\, \mathrm{e}^{\mathrm{i}n(\varphi-
\varphi_0)}.  \\
\end{equation}
Note that except for the radial functions $R_n(r,\tau|r_0)$, all other parts 
of the ansatz are time-independent. 

Observing that for the hard-disk potential, see 
Eq.~\eqref{eq:hard_interaction_potential}, we have that $\mathrm{exp}(-\beta 
U(r)) = \Theta(r-1)$, we can replace the otherwise singular interaction force 
$\partial U(r)/\partial r$ in Eq.~\eqref{eq:EqRelCoord} via
\begin{equation}
\label{eq:DerivativeHeavisidefct}
 - \beta\mathrm{\Theta}(r-1)\, \pdv{U(r)}{r} = \pdv{}{r} \mathrm{e}^{-\beta 
 U(r)} = \mathrm{\delta}(r-1).
\end{equation}
Rewriting Eq.~\eqref{eq:EqRelCoord} for the radial functions 
$R_n(\tau)= R_n(r,\tau|r_0)$, we thus find 
\begin{equation}
\label{eq:DGLrho>d}
\pdv{\tau} R_n(\tau) = \frac{1}{r^2} \left(r \pdv{r} r \pdv{r} - n^2 \right)\,  
R_n(\tau)
\end{equation}
in the domain $r \geq 1$ and for each order $n\in\mathbb{Z}$, and
\begin{equation}
\label{eq:DGLrho=d}
\pdv{r} R_n(\tau) = - \frac{\mathrm{i}n\kappa}{r} R_n(\tau)
\end{equation}
to be satisfied additionally at $r = 1$. Eq.~\eqref{eq:DGLrho=d} can be viewed 
as an extension of an ordinary von Neumann no-flux boundary condition which is 
recovered for $\kappa=0$. This generalized condition can be found in the 
literature under the name of \textit{oblique} boundary conditions, see for 
example~\cite{gilbarg2001elliptic}. Eq.~\eqref{eq:DGLrho>d} is equipped with a 
second boundary condition, which can be found from the natural boundary 
condition on $p(\tau)$, $\lim_{r \to \infty} p(\tau) = 0$, as 
$\lim_{r \to \infty} R_n(\tau) = 0$ and the initial condition on $R_n(\tau)$ 
translates from Eq.~\eqref{eq:initial_cond_inner_p} as $R_n(\tau=0) = 
\delta(r- r_0)/r_0$. Note that Eq.\eqref{eq:DGLrho>d} together with the 
boundary condition \eqref{eq:DGLrho=d} also forms the basis of solving 
the problem by using the language of scattering 
theory~\cite{franosch2010persistent}, the only difference being the introduction 
of the 
the generalized von-Neumann boundary condition for interacting odd-diffusive 
particles.

For the particular solution of Eq.~\eqref{eq:DGLrho>d}, $R_n^\mathrm{part}(\tau)$, 
we make the ansatz 
\begin{equation}
R_n^\mathrm{part}(\tau) = \int \limits_{0}^{\infty} \dd{u}\, w_n(r,u|r_0)\, 
\mathrm{e}^{-\tau u^2},
\end{equation}
which when inserted into Eq.~\eqref{eq:DGLrho>d} shows that $w_n(r,u|r_0)$ 
satisfy the Bessel equation~\cite{abramowitz1968handbook} with solutions 
$J_n(ur)$ and $Y_n(ur)$ as the Bessel functions of first kind and second kind, 
respectively. Matching the particular solution with the initial condition 
$R_n(\tau=0) = \delta(r-r_0)/r_0$ and noting that the delta-distribution 
$\delta(\cdot)$ can be expanded in Bessel functions of the first kind (see 
Eq.~\eqref{eq:IntuJ(au)J(bu)}) we find that $Y_n(ur)$ does not contribute to 
the particular solution. 

After a Laplace transformation with (dimensionless) Laplace variable $s$, the 
homogeneous part of Eq.~\eqref{eq:DGLrho>d} appears as the modified Bessel 
equation~\cite{abramowitz1968handbook} with solutions $I_n(r\sqrt{s})$ and 
$ K_n(r\sqrt{s})$, as the modified Bessel functions of the first and second 
kind, respectively. The Laplace transformation for a function $f(\tau)$ thereby 
is defined as $\mathscr{L}\{f\}(s) = \int_0^\infty\mathrm{d}\tau\ 
\mathrm{exp}(-s\tau)f(\tau)$, noting that $\tau = t/\tau_d$. As the 
homogeneous solution has to satisfy the natural boundary condition on 
$R_n(\tau)$, $I_n(x)$ is no suitable solution as it diverges for $x \to \infty$ 
for every $n\in \mathbb{Z}$ \cite[9.7.1]{abramowitz1968handbook}. We conclude 
that the homogeneous part is solved by $\mathscr{L}\{R_n^\mathrm{hom}\}(s) = 
A_n(s|r_0)\,K_n(r\sqrt{s})$ for some amplitude $A_n(s|r_0)$ to be determined 
by matching with the oblique boundary condition~\eqref{eq:DGLrho=d}.

The relative solution $p(\tau)$ in the Laplace domain thus is given by 
\begin{widetext}
\begin{equation}
\label{eq:pdfWithIntegrals}
\mathscr{L}\{p\}(\mathbf{x},s|\mathbf{x}_0) = \mathrm{\Theta}(r-1) 
\sum_{n = - \infty}^{\infty} \frac{\mathrm{e}^{\mathrm{i}n(\varphi-\varphi_0)}}
{2\mathrm{\pi}} \int \limits_{0}^{\infty} \dd{u}\ \frac{u\, J_n(u r_0)}{s + u^2} 
\left[J_n(u r) -K_n(r\sqrt{s})\, \frac{u\, J^{\prime}_n(u) + \mathrm{i}n\kappa 
\, J_n(u)} { \sqrt{s}\,K^{\prime}_n\left(\sqrt{s}\right) + \mathrm{i}n\kappa \, 
K_n(\sqrt{s})} \right].
\end{equation}
\end{widetext} 
which was already found in Ref.~\cite{kalz2024oscillatory}. There are two types 
of integrals appearing in Eq.~\eqref{eq:pdfWithIntegrals}, which involve 
products of Bessel functions and which we list in 
Appendix~\ref{app:integral_relations}. The integrals $\int_0^\infty\mathrm{d}u\ 
u\, J_n(u r_0)J_n(u b)/(s + u^2)$ can be evaluated using 
Eq.~\eqref{first_bessel_integral}, where $b \in\{1,r\}$. Depending on whether 
$b = r > r_0$, $r = r_0$ or $ r < r_0$, Eq.~\eqref{first_bessel_integral} 
gives different results but we only need to consider the latter two for the 
cases $b=1= r_0$ and $1<r_0$. The second integral $\int_0^\infty\mathrm{d}u\ 
u^2\, J_n(u r_0)J^\prime_n(u)/(s + u^2)$ involves the derivative of a Bessel 
function and therefore is more involved. We list the integral in 
Eq.~\eqref{second_bessel_integral} and again need to differentiate whether 
$r_0 =1$ or $r_0>1$.  Taking into account these different cases, we find for 
the Laplace transformed relative PDF

\begin{widetext}
\begin{equation}
\label{eq:finalSolutionLPD}
\begin{split}
\mathscr{L}\{p\}(\mathbf{x},s|\mathbf{x}_0) = \frac{\Theta(r-1)}{2\pi} 
\sum_{n=-\infty}^{\infty} \mathrm{e}^{\mathrm{i}n(\varphi-\varphi_0)} 
\bigg[ \Theta(r_0-r)\,   I_n(r\sqrt{s})\, K_n(r_0\sqrt{s})  +\Theta(r-r_0)\, 
I_n(r_0\sqrt{s})\, K_n(r\sqrt{s})  \\
- \frac{K_n(r\sqrt{s})\, K_n(r_0\sqrt{s})}{\sqrt{s} \, K_n^\prime(\sqrt{s}) 
+ \mathrm{i}n\kappa \, K_n(\sqrt{s})}\left(\sqrt{s} \, I_n^\prime(\sqrt{s}) 
+ \mathrm{i} n \kappa\, I_n(\sqrt{s}) - \frac{\delta(r_0-1)}{2K_n(\sqrt{s})}\right) 
\bigg].
\end{split}
\end{equation}
\end{widetext}
We can Laplace invert some parts of Eq.~\eqref{eq:finalSolutionLPD} as
\begin{align}
\label{eq:LT_InKn}
    \mathscr{L}^{-1} \{  I_{n}(r_0\sqrt{s})\,& K_{n}(r \sqrt{s})  \} = 
    \mathscr{L}^{-1} \left\{  I_{n}(r\sqrt{s})\, K_{n}(r_0 \sqrt{s})  \right\} 
    \nonumber\\
    &= \frac{\exp \left(-\frac{r_0^2+r^2}{4\tau}\right)}{2 \tau} I_{n}
    \left(\frac{r_0r}{2\tau} \right),
\end{align}
\cite[13.96]{oberhettinger1973tables}. Together with decomposing 
$\mathrm{e}^{\mathrm{i}x} = \cos{x} + \mathrm{i} \sin{x}$ in 
Eq.~\eqref{eq:finalSolutionLPD}, this gives the Fourier expanded relative PDF as
\begin{align}
\label{eq:finalSolTimeDomain}
    &p(\mathbf{x}, \tau|\mathbf{x}^0) =\frac{\Theta(r-1)}{2\pi} 
    \Bigg[ a_0(r,\tau|r_0)  \nonumber \\
    &+ 2 \sum_{n=1}^{\infty}  \begin{pmatrix} a_n(r,\tau|r_0) \\ b_n(r,\tau|r_0) 
    \end{pmatrix}  \cdot  \begin{pmatrix} \cos\left(n(\varphi-\varphi_0)\right)\\  
        \sin\left(n(\varphi-\varphi_0)\right) \end{pmatrix} \Bigg],
\end{align}
where $a_n$ and $b_n$ are given in Eqs.~\eqref{eq:fourier_coefficient_a} 
and~\eqref{eq:fourier_coefficient_b} in the main text. Note that by using the 
symmetry in the order of the modified Bessel functions, $I_{-n}(x) = I_{n}(x)$ 
and $K_{-n}(x) = K_{n}(x)$  \cite[9.6.6]{abramowitz1968handbook}, we can 
restrict the sum on positive modes only.

\section{Integral expressions}
\label{app:integral_relations}
To evaluate the integrals in Eq.~\eqref{eq:pdfWithIntegrals}, we use the 
following tabulated integrals, listed in this Appendix.

Following Arfken and Weber's 
\textit{Mathematical Methods for Physicists}~\cite{arfken2005mathematical}, we 
find that Bessel functions of (integer) order $n$, $J_n$, obey the integral 
relation
\begin{equation}
\label{eq:IntuJ(au)J(bu)}
    \int \limits_{0}^{\infty} \dd{u}\, u\, J_{n}(a u)\, J_{n}(b u) = 
    \frac{\mathrm{\delta}(a-b)}{a},
\end{equation}
valid for $n>-1/2$ and $a,b$ some real-valued constants 
\cite[11.59]{arfken2005mathematical}. 

To evaluate the remaining integrals of Bessel-functions, we draw on Gradshteyn 
and Ryzhik's \textit{Table of Integrals, Series 
and Products}~\cite{gradshteyn2007table} and Abramowitz and Stegun's 
\textit{Handbook of Mathematical Functions}~\cite{abramowitz1968handbook}. The 
first relevant integral is

\begin{equation}
\label{first_bessel_integral}
     \int \limits_{0}^{\infty} \dd{u}\ \frac{u\ J_n(au) J_n(bu)}{u^2 + c^2} 
     = \begin{cases}I_n(ac) \, K_n(bc), & 0<a<b \\ I_n(ac) \, K_n(ac), & 0<a=b 
        \\ I_n(bc) \, K_n(ac), & 0<b<a\end{cases},
\end{equation}
which is valid for $n >-1$ and $c>0$ \cite[6.541, 6.535]{gradshteyn2007table}. 
Here $ I_n$, $K_n$ are the modified Bessel functions of the first and second 
kind, and $a,b,c$ some (real-valued) constants. Together with the symmetry 
relations for the order of Bessel functions, $J_{-n}(x) = (-1)^n J_n(x)$, 
$I_{-n}(x) = I_{n}(x)$ and $K_{-n}(x) = K_{n}(x)$ 
\cite[9.1.5, 9.6.6]{abramowitz1968handbook}, Eq.~\eqref{first_bessel_integral} 
can be further extended to the relevant cases of $n<0$. 

The second relevant integral is $\int_{0}^{\infty} \dd{u}\ u^2\, J_{n}(au)\,  
J_{n}^\prime(bu)/(u^2 + c^2)$, for some real-valued, positive constants $a,b,c$. 
Using $\mathrm{d}J_{n}(x)/\mathrm{d}x = (J_{n-1}(x) - J_{n+1}(x))/2$ 
\cite[8.471]{gradshteyn2007table}, the integral can be evaluated as 
\begin{align}
\label{intermediate_bessel_integral}
     &\int \limits_{0}^{\infty} \dd{u}\ \frac{u^2\ J_{n_1}(au) J_{n_2}(bu)}
     {u^2 + c^2} \nonumber \\
     &=  \begin{cases}  (-1)^{\alpha_+}\, c\ I_{n_1}(ac)\, K_{n_2}(bc), & 0<a<b \\
     (-1)^{\alpha_-}\, c\ I_{n_2}(bc)\, K_{n_1}(ac), & 0<b<a\end{cases},
\end{align}
which is valid for $n_2 >-1$ \cite[6.577]{gradshteyn2007table}. The exponents 
are $\alpha_{\pm} = (1 \pm (n_1 - n_2))/2 \in \mathbb{N}_0$. We use 
Eq.~\eqref{intermediate_bessel_integral} for the case of $n_1 = n\pm 1$ and 
$n_2=n$. By relying on the symmetry relations for the orders of the Bessel 
functions, we can extend Eq.~\eqref{intermediate_bessel_integral} again to the 
negative cases of $n_1 = -(n\pm 1)$ and $n_2=-n$. Thus, together with 
\cite[8.468]{gradshteyn2007table}, we find 
\begin{equation}
\label{almost_second_bessel_integral}
     \int \limits_{0}^{\infty} \dd{u}\ \frac{u^2\ J_{n}(au) J_{n}^\prime(bu)}
     {u^2 + c^2} = \begin{cases} c\ K_{n}(bc)\, I_{n}^\prime(ac), 
 & 0<a<b \\ \frac{n}{b} K_{n}(ac)\, I_{n}(bc), & 0<b<a\end{cases},
\end{equation}
valid for all $n \in \mathbb{Z}$, but limited to $b\neq a$. 

In order to generalize Eq.~\eqref{almost_second_bessel_integral} to the case of 
$b=a >0$, which can not be found in Ref.~\cite{gradshteyn2007table}, we 
generalize the relation, which was already derived in 
Ref.~\cite{kalz2024oscillatory} for $n=1$ to an arbitrary $n \in \mathbb{Z}$. 
Therefore, by using the indefinite integral $\int\dd{u}\ J_{n}(u) 
J_{n}^\prime(u) = J_{n}^2(u)/2 + const$, we can partially integrate and find an 
integral of the form of Eq.~\eqref{first_bessel_integral}, which, using the 
Wronskian $\mathcal{W}[I_{n}(x), K_{n}(x)] = I_{n}(x)\, K_{n+1}(x) + I_{n+1}(x)\, 
K_{n}(x)= 1/x$ \cite[9.1.15]{abramowitz1968handbook}, gives
\begin{align}
\label{second_bessel_integral}
&\int \limits_{0}^{\infty} \dd{u}\, \frac{u^2\ J_{n}(au) J_{n}^\prime(au)}
{u^2 + c^2} \nonumber\\
&= \begin{cases} c\ K_{n}(bc)\, I_{n}^\prime(ac), &0<a<b \\ c\ K_{n}(ac)\, 
I_{n}^\prime(ac) - \frac{1}{2a}, & 0<a=b \\ \frac{n}{b} K_{n}(ac)\, I_{n}(bc),
& 0<b<a\end{cases}.
\end{align}

To prove the normalization of the zeroth order mode $\int\mathrm{d}\mathbf{x}\ 
p_0 =1$, Eq.~\eqref{eq:mean_positional_PDF} in the main text, we transform the 
integrals over $I_0$ and $K_0$ into higher order Bessel functions evaluated at 
the boundaries $r=1$ and $r \to \infty$ by using the recursion relations 
\cite[8.486]{gradshteyn2007table}
\begin{equation}
\left(\frac{1}{u}\ \dv{u} \right)^m  \left( u^n\ I_{n}(u) \right) = 
u^{n - m}\ I_{n-m}(u), 
\end{equation}
and 
\begin{equation}
\left(\frac{1}{u}\ \dv{u} \right)^m  \left( u^n\ K_{n}(u) \right) = 
(-1)^m\ u^{n - m}\ K_{n-m}(u),
\end{equation}
which specifically imply for $m = 1$ and $n= 1$ to $u\, I_0(u) = 
\mathrm{d}/\mathrm{d}u\, (u\, I_1(u))$ and $u\, K_0(u) = -\mathrm{d}/
\mathrm{d}u\, (u\, K_1(u))$. The integrals over $I_0(u)$ and $K_0(u)$ thus turn 
into evaluating $u\, I_1(u)$ and $u\, K_1(u)$ at the integration bounds and the 
upper bound vanishes by the asymptotic expansion of $\lim_{u\to \infty} uK_n(u) 
\sim \lim_{u\to \infty} u\, \exp(-u) = 0$ \cite[8.451]{gradshteyn2007table}. 
The remaining expressions from the lower bounds are evaluated by using the 
Wronskian $\mathcal{W}[I_{n}(x), K_{n}(x)]$, to be exactly 1.


\begin{thebibliography}{98}
\bibitem [{tri()}]{trick}%
    \BibitemOpen
    \href@noop {} {}\bibinfo {note} {For the hard-disk potential $U(r)=0$ for
    $r>1$ (rescaled space variable) and $U(r)=\infty$ otherwise we observe that
    $\mathrm{exp}(-\beta U(r))=\Theta(r-1)$. We thus find on the one hand that
    $\nabla \mathrm{exp}(-\beta U(r)) = -\beta\, \mathrm{exp}(-\beta U(r))\,
    \nabla U(r)) = \beta\, \Theta(r-1)\, \mathbf{F}(r)$, where $\mathbf{F}(r) =
    -\nabla U(r)$ is the interaction force and on the other hand that $\nabla
    \mathrm{exp}(-\beta U(r)) =\delta(r-1)\, \mathbf{e}(\varphi)$, where
    $\mathbf{e}(\varphi) = (\cos(\varphi), \sin(\varphi))^\mathrm{T}$ is the
    radial unit vector.}\BibitemShut {Stop}%
\end{thebibliography}

%

\end{document}